\documentclass{iaus}
\usepackage{graphicx}

\title[Gas inflows and metallicity evolution in galaxy pairs] 
{Gas inflows, star formation and metallicity evolution in galaxy pairs}

\author[P. Di Matteo et al.]   
{Paola Di Matteo$^1$,
 Marco Montuori$^2$,
 Matthew D. Lehnert$^1$,
 Fran\c coise Combes$^3$ 
 \and Benoit Semelin$^3$}

\affiliation{$^1$ GEPI, Observatoire de Paris, CNRS, Universit\'e Paris Diderot \\ 5, Place Jules Janssen, 92190 Meudon, France \\ email: {\tt paola.dimatteo@obspm.fr, matthew.lehnert@obspm.fr} \\[\affilskip]
$^2$SMC-ISC-CNR \& Physics Department, Universit\'a di Roma La Sapienza \\Pl. Aldo Moro 2, 00185 Rome, Italy \\email: {\tt montuorm@roma1.infn.it }\\[\affilskip]
$^3$Observatoire de Paris, LERMA, CNRS, UPMC\\61 Avenue de l'Observatoire, 75014 Paris, France\\{\tt francoise.combes@obspm.fr, benoit.semelin@obspm.fr}}

\pubyear{2008}
\volume{xxx}  
\pagerange{119--126}
\jname{Title of your IAU Symposium}
\editors{A.C. Editor, B.D. Editor \& C.E. Editor, eds.}
\begin{document}

\maketitle

\begin{abstract}
It has been known since many decades that galaxy interactions can induce star formation (hereafter SF) enhancements and that one of the driving mechanisms of this enhancement is related to gas inflows into the central galaxy regions, induced by asymmetries in the stellar component, like bars. In the last years many evidences have been accumulating, showing that interacting pairs have central gas-phase metallicities lower than those of field galaxies, by $\sim$ 0.2-0.3 dex on average. These diluted ISM metallicities have been explained as the result of inflows of metal-poor gas from the outer disk to the galaxy central regions. A number of questions arises: What's the timing and the duration of this dilution? How and when does the SF induced by the gas inflow enrich the circum-nuclear gas with re-processed material? Is there any correlation between the timing and strength of the dilution and the timing and intensity of the SF? By means of Tree-SPH simulations of galaxy major interactions, we have studied the effect that gas inflows have on the ISM dilution, and the effect that the induced SF has, subsequently, in re-enriching the nuclear gas. In this contribution, we present the main results of this study. 
\end{abstract}

\firstsection 
\section{Introduction}

In the local Universe, the most luminous galaxies are associated
to gas-rich interactions and mergers (\cite[Sanders \& Mirabel (1996)]{sanmir96}), but this is not reciprocal. There is increasing evidence that galaxy interactions and mergers are not sufficient conditions to drive strong starbursts (\cite[Bergvall et al. (2003)]{berg03},  \cite[Knapen \& James (2009)]{knap09},  \cite[Ellison et al. (2011)]{ell11} in this volume, \cite[Jogee et al. (2009)]{jogee09} and  \cite[Robaina et al. (2009)]{robaina09}).

In fact, the amplitude of the SF enhancement depends on a number of
parameters: morphology, gas content, strength of the tidal
effects, mass ratio... as numerical models show. Both the timing and the strength of the SF burst depends, for example, on the morphology of the interacting galaxies: the presence of a bulge can stabilise galaxy disks against bar formation, limiting the amplitude of the SF enhancement at the first close passage, so that most of the gas is not consumed till the final phases of the interaction, when a strong burst can occur (\cite[Mihos \& Hernquist (1994)]{mihos94}). For a given morphology, the characteristics of the orbit, as well as the strength of the tidal effects, can play a major role in determining the amplitude of the SF increase (see \cite[Di Matteo et al. (2007)]{dimatteo07}, \cite[Di Matteo et al. (2008)]{dimatteo08}): in general, there is a negative correlation between the amplitude of the SF burst in the final phases of coalescence and the tidal forces exerted at the first close passage, which is due to the large amount of gas dragged outside the galaxy by tidal tails in strong interactions.
The strongest SF enhancements are generally associated to major mergers, and the SF efficiency decreases with the mass ratio (\cite[Cox et al. (2008)]{cox08}). Whereas numerical simulations usually predicts that the strongest SF enhancements are localized in the central galaxy regions, one has to take into account that the spatial distribution of the star forming regions depends also on the SF prescription adopted in the models: a more extended SF is indeed predicted if the local rate of energy dissipation in shocks is included (\cite[Barnes (2004)]{barnes04}). Recently \cite[Teyssier et al. (2010)]{teyssier10} have also pointed out that both the intensity and the distribution of SF depends on the numerical resolution adopted in the simulations. This is a point that will surely deserve further investigation in the future.

The physical mechanism responsible of the SF enhancement in the central regions of interacting galaxies has been described in detail by \cite[Mihos \& Hernquist (1996)]{mihos96}: during an interaction, galaxy disks are destabilized and stellar asymmetries like bars can be produced (or strenghtened). These asymmetries constitute an efficient mechanism to remove angular momentum from the gaseous component present in the disk. Losing its rotational support, the gas falls into the central region, where the SF enhancement is produced.\\
Thus the question is to understand if there is any way to trace these gas inflows during galaxy interactions, and what kind of signatures they leave in the galaxy disk.

Recently, \cite[Kewley et al. (2006)]{kewley06} have derived the luminosity-metallicity relation for a sample of local galaxy pairs and compared it with that of nearby field galaxies. They found that pairs with small projected separations ( $s < 20~{\rm kpc}~h^{-1}$) have systematically lower metallicities than either isolated galaxies or pairs with larger separations, for a given luminosity. They also found a correlation between gas metallicity and burst strengths - all galaxies in their interacting sample with strong central bursts having close companions and metallicities lower than the comparable field galaxies or pairs with wider separations. \cite[Rupke et al. (2008)]{rupke08}, in a study of local ULIRGs and LIRGs, found that the metal abundance in these intense starbursts is a factor of two lower than that of galaxies of comparable luminosity and mass. Also the radial gas metallicity gradients seem to be affected (flattened) by the interaction with a companion, as shown by \cite[Rupke et al. (2010)]{rupke10}. All these results have been generally explained by the following arguments: if, prior to the interaction, the gas is distributed in the disk with a negative metallicity 
gradient, with the outer regions being relatively more metal poor than the inner ones as  observed in local galaxies, then part of this low-metallicity gas will fall into the nuclear  regions, diluting the pre-existing gas with gas of lower metallicity. 
Of course, this dilution will last  until the SF, which is enhanced by the gas inflow, releases reprocessed metals to enrich the gas.

\section{The simulations}
To study the cycle of gas metallicity dilution, and subsequent enrichment due to SF, we analyzed 96 simulations of major (mass ratio 1:1)  mergers and flybys, realized in the framework of the GalMer Project (see \cite[Chilingarian et al. (2010)]{chili10} and \emph{http://galmer.obspm.fr}). SF and metal enrichment from SNe are included in our models. A detailed description of the numerical techniques adopted and of the initial galaxy models is given in \cite[Montuori et al. (2010)]{montuori10}. In the following, we present  the main results of this study (see \cite[Montuori et al. (2010)]{montuori10} for a complete discussion).

\section{Results}
\subsection{Gas flows: where and when?}
The cycle of dilution and enrichment, discussed at the end of Sect.1, is shown in Fig. 1  for some of our galaxy merger simulations. We have defined the metallicity dilution as  $z/z_{\rm iso}< 1$, where $z$ is the metallicity of a galaxy in the pair during the interaction and  $z_{\rm iso}$ is the corresponding value for the same galaxy evolved isolated.
We find that as soon as the two galaxies have undergone their first pericentre passage, an intense inflow of gas takes place, fueling the SF enhancement. The time of strong enhancement in the SFR also coincides with the time of high-dilution. Typically, we find that the dilution lasts for 2 $\times  10^9$ years, with half of the sample substaining a $z/z_{\rm iso}\le 0.8$ for less than 5 $\times 10^8$ yrs. We emphasize that the dilution occurs also in flybys (pairs that interact but do not merge): in this case, it is still visible far after the pericentre passage, when the separation between the two galaxies is several hundreds of kpc. 

\begin{figure}[b]
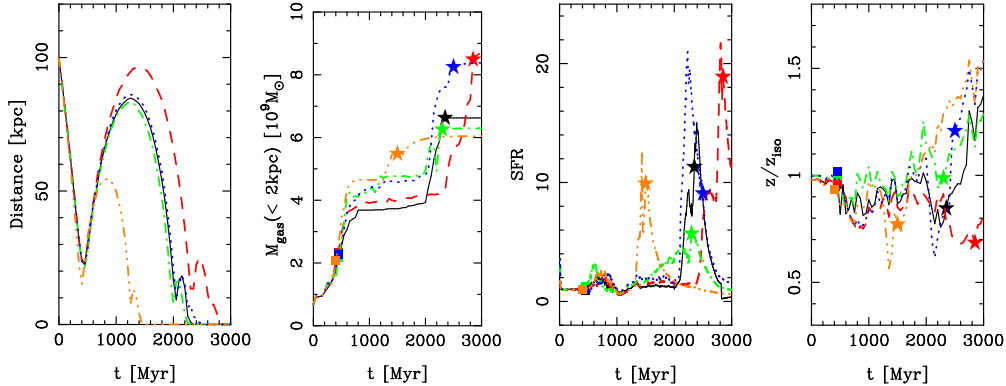

\begin{center}
 \includegraphics[width=2.in,angle=270]{14304fg1.ps} 
 \includegraphics[width=2.in,angle=270]{14304fg2.ps} 
 \includegraphics[width=2.in,angle=270]{14304fg3.ps} 
 \includegraphics[width=2.in,angle=270]{14304fg4.ps} 
 \caption{(First panel): relative distance of the two galaxy centres versus time for some simulated orbits. (Second panel): gas mass inside 2 kpc from one of the two galaxy centres versus time. (Third panel): evolution of the star formation rate (SFR) relative to the rate for the corresponding initial galaxies evolved in isolation. (Fourth panel): evolution of the central gas metallicity. The metallicity has been normalised to that of the initial galaxies used in the merger simulations evolved isolated. In the second, third, and fourth panel, squares and asterisks respresent respectively the time of the first pericentre passage and the time of final coalescence for the different orbits. From Montuori et al. (2010).}
   \label{fig1}
\end{center}
\end{figure}

\begin{figure}[t]
\begin{center}
 \includegraphics[width=2.5in,angle=0]{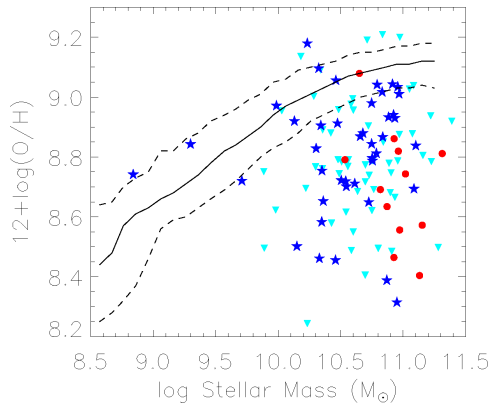} 
 \caption{Monte-carlo simulation of the metallicity distribution predicted by the models compared to the data from Rupke et al. (2008). The up-side down triangles are the simulated results, the blue stars and red circle are galaxy values taken directly from Rupke et al. (2008) as described there. From Montuori et al. (2010).}
   \label{fig4}
\end{center}
\end{figure}
\subsection{Star formation and metal dilution}
The dilution of the gas metallicity takes place at approximately the same time as the strongest enhancement in the SF. 
Before the peak in the SF rate, galaxies tend to populate a region of low to moderate SFRs and dilutions (typically $0.7 \le z/z_{\rm iso}\le 1$). During the burst phase and peak of SF rate, SNe start to enrich the gas, and indeed, once the burst is over, galaxies have low SF rates and high-metallicities. The first two phases of the relation between SF and dilution are associated with before, during, and after the first pericentre passage and then close to final coalescence of the merging galaxies, the last phase in the metallicity evolution is always associated with well after their first pericentre passage and is near or after coalescence. We refer the reader to \cite[Montuori et al. (2010)]{montuori10}, and in particular to Fig.4 in that paper, for a more detailed discussion.

\subsection{Models versus observations}
Our models can explain the full range of circumnuclear metallicities observed in violent mergers in the local (Rupke et al. 2008) or distant universe (Rodrigues et al. 2008). By violent mergers, we mean  galaxies which are luminous in the infrared and have SF rates several to almost 100 times that of the Milky Way. To show that our models can reproduce the data, we have made a monte-carlo simulation of the metallicity distribution we might obtain if we observed a set of galaxies near their maximum SF rate and allow for a range in their properties and the intrinsic scatter in the mass-metallicity relationship. The results of this comparison are shown in Fig.2, demonstrating that strong gas inflows, and subsequent strong SF enhancements, are responsible of the strong dilution observed in local LIRGs and ULIRGs. The models can indeed reproduce both the average value of this dilution (0.2-0.3 dex) and its dispersion.


\begin{thebibliography}{}

\bibitem[Barnes (2004)]{barnes04}{Barnes, J.˜E.} 2004, \textit{MNRAS}, 350, 798
\bibitem[Bergvall \etal\ (2003)]{berg03}{Bergvall, N., Laurikainen, E., Aalto, S.} 2003, \textit{A\&A}, 405, 31
\bibitem[Chilingarian \etal\ (2010)]{chili10}{Chilingarian, I.˜V., Di Matteo, P., Combes, F., Melchior, A.-L., Semelin, B.} 2010, \textit{A\&A}, 518, 61
\bibitem[Cox \etal\ (2008)]{cox08}{Cox, T.˜J., Jonsson, P., Somerville, R.˜S., Primack, J.˜R., Dekel, A.} 2008, \textit{MNRAS}, 384, 386
\bibitem[Di Matteo \etal\ (2007)]{dimatteo07}{Di Matteo, P., Combes, F., Melchior, A.-L., Semelin, B.} 2007, \textit{A\&A}, 468, 61
\bibitem[Di Matteo \etal\ (2008)]{dimatteo08}{Di Matteo, P., Bournaud, F., Martig, M., Combes, F., Melchior, A.-L., Semelin, B.} 2008, \textit{A\&A}, 492, 31
\bibitem[Ellison \etal\ (2011)]{ell11}{Ellison, S., Patton, D.˜R., Nair, P., Simard, L., Mendel, J.˜T., McConnachie, A.˜W., Scudder, J.˜M.} 2011, \textit{this volume}; astro-ph/1101.3566
\bibitem[Jogee \etal\ (2009)]{jogee09}{Jogee, S., Miller, S.˜H., Penner, K., Skelton, R.˜E., Conselice, C.˜J. et al.} 2009, \textit{ApJ}, 697, 1971
\bibitem[Kewley \etal\ (2006)]{kewley06}{Kewley, L.˜J., Geller, M.˜J., Barton, E.˜J.} 2006, \textit{AJ}, 131, 2004
\bibitem[Knapen \& James(2009)]{knap09}{Knapen, J.˜H. \& James, P.˜A.} 2009, \textit{ApJ}, 698, 1437
\bibitem[Mihos \& Hernquist(1994)]{mihos94}{Mihos, J.˜C. \& Hernquist, L.} 1994, \textit{ApJL}, 431, 9
\bibitem[Mihos \& Hernquist(1996)]{mihos96}{Mihos, J.˜C. \& Hernquist, L.} 1996, \textit{ApJ}, 464, 641
\bibitem[Montuori \etal\ (2010)]{montuori10}{Montuori, M., Di Matteo, P., Lehnert, M.˜D., Combes, F., Semelin, B.} 2010, \textit{A\&A}, 518, 56
\bibitem[Robaina \etal\ (2009)]{robaina09}{Robaina, A.˜R., Bell, E.˜F., Skelton, R.˜ E., McIntosh, D.˜H., Somerville, R.˜S. et al.} 2009, \textit{ApJ}, 704, 324
\bibitem[Rodrigues \etal\ (2008)]{rodrig08}{Rodrigues, M., Hammer, F., Flores, H., Puech, M., Liang, Y.˜C. et al.} 2008, \textit{A\&A}, 492, 371
\bibitem[Rupke \etal\ (2008)]{rupke08}{Rupke, D.˜S.˜N., Veilleux, S., Baker, A.˜J.} 2008, \textit{ApJ}, 674, 172
\bibitem[Rupke \etal\ (2010)]{rupke10}{Rupke, D.˜S.˜N., Kewley, L.˜J., Chien, L.-H.} 2010, \textit{ApJ}, 723, 1255
\bibitem[Sanders \& Mirabel(1996)]{sanmir96}{Sanders, D.˜B. \& Mirabel, I.˜F.} 1996, \textit{ARA\&A}, 34, 749
\bibitem[Teyssier \etal\ (2010)]{teyssier10}{Teyssier, R., Chapon, D., Bournaud, F.} 2010, \textit{ApJL}, 720, 149



\end{thebibliography}
\end{document}